\documentclass[10pt,twocolumn,preprintnumbers,amsmath,amssymb,nofootinbib
,superscriptaddress]{revtex4-2}
\usepackage{amsmath,amssymb,amsthm,mathrsfs}
\usepackage{epsfig}
\usepackage{graphicx}
\usepackage[usenames,dvipsnames]{color}
\usepackage{subfigure}
\usepackage{slashed}
\usepackage{comment}
\usepackage[normalem]{ulem}
\usepackage{tikz}
\usepackage[compat=1.1.0]{tikz-feynman}
\usepackage[colorlinks,citecolor=blue]{hyperref}
\usepackage{color}

\usepackage{multirow}
\usepackage{tikz-feynman}
\tikzfeynmanset{compat=1.1.0}

\usepackage{tikzsymbols}

\usepackage{xcolor}
\usepackage{soul}

\begin{document}
\title{Linear dynamics and classical tests of the gravitational quantum field theory}
\author{Yuan-Kun~Gao\footnote{gaoyuankun17@mails.ucas.ac.cn}}
\affiliation{International Centre for Theoretical Physics Asia-Pacific (ICTP-AP), University of Chinese Academy of Sciences (UCAS), Beijing 100190, China}
\affiliation{Taiji Laboratory for Gravitational Wave Universe (Beijing/Hangzhou), UCAS, Beijing 100049, China}

\author{Da~Huang\footnote{dahuang@bao.ac.cn}}
\affiliation{National Astronomical Observatories, Chinese Academy of Sciences, Beijing 100101, China}
\affiliation{School of Fundamental Physics and Mathematical Sciences, Hangzhou Institute for Advanced Study, UCAS, Hangzhou 310024, China}
\affiliation{International Centre for Theoretical Physics Asia-Pacific (ICTP-AP), University of Chinese Academy of Sciences (UCAS), Beijing 100190, China}
\affiliation{Taiji Laboratory for Gravitational Wave Universe (Beijing/Hangzhou), UCAS, Beijing 100049, China}

\author{Yong-Liang~Ma\footnote{ylma@ucas.ac.cn}}
\affiliation{Nanjing University, Suzhou, 215163, China}
\affiliation{School of Fundamental Physics and Mathematical Sciences, Hangzhou Institute for Advanced Study, UCAS, Hangzhou 310024, China}
\affiliation{Taiji Laboratory for Gravitational Wave Universe (Beijing/Hangzhou), UCAS, Beijing 100049, China}
\affiliation{International Centre for Theoretical Physics Asia-Pacific (ICTP-AP), University of Chinese Academy of Sciences (UCAS), Beijing 100190, China}

\author{Yong~Tang\footnote{tangy@ucas.ac.cn}}
\affiliation{School of Astronomy and Space Sciences, UCAS, Beijing 100049, China}
\affiliation{School of Fundamental Physics and Mathematical Sciences, Hangzhou Institute for Advanced Study, UCAS, Hangzhou 310024, China}
\affiliation{International Centre for Theoretical Physics Asia-Pacific (ICTP-AP), University of Chinese Academy of Sciences (UCAS), Beijing 100190, China}
\affiliation{Taiji Laboratory for Gravitational Wave Universe (Beijing/Hangzhou), UCAS, Beijing 100049, China}

\author{Yue-Liang~Wu\footnote{ylwu@ucas.ac.cn}}
\affiliation{International Centre for Theoretical Physics Asia-Pacific (ICTP-AP), University of Chinese Academy of Sciences (UCAS), Beijing 100190, China}
\affiliation{Taiji Laboratory for Gravitational Wave Universe (Beijing/Hangzhou), UCAS, Beijing 100049, China}
\affiliation{School of Fundamental Physics and Mathematical Sciences, Hangzhou Institute for Advanced Study, UCAS, Hangzhou 310024, China}
\affiliation{CAS key laboratory of theoretical Physics, Institute of Theoretical Physics, Chinese Academy of Sciences, Beijing 100190, China}

\author{Yu-Feng~Zhou\footnote{yfzhou@itp.ac.cn}}
\affiliation{CAS key laboratory of theoretical Physics, Institute of
	Theoretical Physics, Chinese Academy of Sciences, Beijing 100190,
	China}
\affiliation{School of Physics, UCAS, Beijing 100049, China}
\affiliation{School of Fundamental Physics and Mathematical Sciences, Hangzhou
	Institute for Advanced Study, UCAS, Hangzhou 310024, China}
\affiliation{International Centre for Theoretical Physics Asia-Pacific (ICTP-AP), University of Chinese Academy of Sciences (UCAS), Beijing 100190, China}

\date{\today}
\begin{abstract}
\noindent 
We explore the new physics phenomena of gravidynamics governed by the inhomogeneous spin gauge symmetry based on the gravitational quantum field theory. Such a gravidynamics enables us to derive the generalized Einstein equation and an equation beyond it. To simplify the analyses, we linearize the dynamic equations of gravitational interaction by keeping terms up to the leading order in the dual gravigauge field. We then apply the linearized dynamic equations into two particular gravitational phenomena. Firstly, we consider the linearized equations in the absence of source fields, which is shown to have five physical propagating polarizations as gravitational waves, i.e., two tensor modes, two vector modes, and one scalar, instead of two tensor polarizations in the general relativity. Secondly, we examine the Newtonian limit in which the gravitational fields and the matter source distribution are weak and static. By deriving the associated Poisson equation, we obtain the exact relation of the fundamental interaction coupling in the gravidynamics with the experimentally measured Newtonian constant. We also make use of non-relativistic objects and relativistic photons to probe the Newtonian field configurations. In particular, the experiments from the gravitational deflection of light rays and the Shapiro time delay can place stringent constraints on the linearized gravidynamics in the gravitational quantum field theory.
\end{abstract}

\maketitle

\section{Introduction}\label{s1}
Among the fundamental interactions, the gravitation is the weakest and the most mysterious. The standard theory of gravitation at present is Einstein's General Relativity (GR)~\cite{Einstein:1915ca, Einstein:1916vd}, in which the gravity is described by the Riemannian geometry of curved spacetime. Since its birth in 1915, the GR has withstood all the astrophysical and cosmological tests, which include four classical ones: (i) the gravitational redshift of photons~\cite{Pound:1960zz,Pound:1965zz,Vessot:1980zz}; (ii) the deflection of light~\cite{Dyson:1920cwa,Shapiro:2004zz,Lambert:2009xy,lambert2011improved}; (iii) the perihelion advance of Mercury~\cite{Einstein:1915bz}; and (iv) the time delay of light~\cite{Shapiro:1964uw,Bertotti:2003rm} (for a recent review on the classical tests of GR, see e.g. Ref.~\cite{Will:2014kxa} and references therein). The discovery of the binary pulsar B1913+16 by Hulse and Taylor in 1974~\cite{Hulse:1974eb} has verified the GR by providing the first evidence of the existence of gravitational radiation~\cite{Taylor:1979zz,1980NYASA.336..442T}. In 2015, the first direct observation by advanced LIGO observatories of the gravitational wave (GW) signal emitted from the merger of binary black holes~\cite{LIGOScientific:2016aoc,LIGOScientific:2016lio} has provided a further support to the GR.  

On the other hand, the electromagnetic, weak and strong interactions are all described by quantum field theories (QFTs) based on the gauge principle~\cite{Yang:1954ek}. Thus, it is tempting to write down the gravity in terms of the gauge language, and numerous efforts were explored in early studies (for review articles  see e.g. Refs.~\cite{Hehl:1976kj,Ivanenko:1983fts,Hehl:1994ue}. Whereas most gravity gauge theories were formulated based on Riemannian geometry on curved space-time or Poincar\'e group in coordinate space-time rather than in Hilbert space of fields. 
Recently, a gauge theory of gravity has been constructed in the framework of gravitational quantum field theory (GQFT)~\cite{Wu:2022mzr} and unified with the standard model to build the hyperunified field theory(see Ref.~\cite{Wu:2022aet} for the latest review of the GQFT and the hyperunified field theory as well as the references therein). It has been shown that the gravitational force and the spin gauge force are governed by the inhomogeneous spin gauge symmetry ${\rm WS}(1,3) = {\rm SP}(1,3)\rtimes {\rm W}^{1,3}$, in which the spin gauge field corresponding to the spin gauge symmetry ${\rm SP}(1,3)$ is introduced to characterize the spin gauge interaction and the gravigauge field, while the translation-like ${\cal W}_e$-spin gauge symmetry ${\rm W}^{1,3}$ in Hilbert space of spinor fields emerges to describe the gravitational force. A biframe spacetime appears to reveal the nature of spacetime in which the globally flat Minkowski spacetime plays the role of the base spacetime while the locally flat gravigauge spacetime acts as a fiber. As a result, the gravitational dynamics is described in terms of the gravigauge and spin gauge fields. Phenomenologically, there have already been many studies in the framework of GQFT~\cite{Wu:2022aet}, including discussions on inflation~\cite{Wang:2023hsb}, dark matter, and particle physics.

Note that the gravitational dynamics in the GQFT is highly nonlinear~\cite{Wu:2022mzr,Wu:2022aet}, which makes its application rather difficult. However, when the gravitational effects are weak, the gravitational dynamics becomes easy and tractable, so that we can expand the equations of motion in terms of perturbations at the linear level.  
In the present work, we would like to explore the gravitational physics in this linearized GQFT. We start by deriving the linearized gravitational field equations, which are then applied to two special situations. Firstly, we shall consider the free gravitational equations without any matter sources, aiming to investigate how many and what physical polarizations are contained in the propagating GWs. Next, we turn to the Newtonian limit, in which the gravitational field and matter sources are weak and static. By solving the linearized equations, we can obtain the well-known Poisson equation which governs the associated Newtonian potential. 
We will also make use of both non-relativistic and relativistic test bodies to probe this Newtonian configuration. In particular, when the probing particles are photons, we can constrain the GQFT with existing experiments, such as the deflection of a light ray and the Shapiro time delay. 

The paper is organized as follows. In Sec.~\ref{SecGT}, we shall derive the linearized gravitational equations of gravigauge fields in the GQFT. We shall apply in Sec.~\ref{SecFree} the obtained equations to the free field case, and examine the physical degrees contained in the propagating GWs. In Sec.~\ref{SecNew}, we turn to the Newtonian limit, which is tested by the non-relativistic test bodies and relativistic photons. Finally, we conclude and comment in Sec.~\ref{SecSum}.

\section{Gravitational Equations of Gravigauge Field at the Linear Level}\label{SecGT}

In this section, we linearize the gravitational equations of gravigauge fields in the 
GQFT~\cite{Wu:2022mzr}. Our derivation begins with the gauge-type formulation of the gravidynamics given in Eq.~(159) of Ref.~\cite{Wu:2022mzr}
\begin{eqnarray}\label{EqGT}
	\partial_\nu \tilde{F}_a^{\,\mu\nu} = J_a^{\,\mu}\,,
\end{eqnarray}
where the definition of $\tilde{F}_a^{\,\mu\nu}$ and various contributions to the source $J_a^{\,\mu}$ are defined in Ref.~\cite{Wu:2022mzr}. Note that the basic ingredient in the gravidynamics is the gravigauge field $\chi_{\,\mu}^a$, which can always be written as follows:
\begin{eqnarray}\label{Gpert}
	\chi_{\,\mu}^a \equiv \eta^a_\mu + \frac{1}{2} h^a_{\,\mu}\,,
\end{eqnarray} 
where $\eta^a_\mu$ is regarded as the background field while $h^a_{\,\mu}$ is the redefinition of the field variable $\chi^a_{\,\mu}$. If we further require that $h^a_\mu$ is a weak 
perturbation, 
we can expand the dual gravigauge field as $\hat{\chi}^{\,\mu}_a = (\eta^a_\mu + \frac{1}{2} h^a_{\,\mu} )^{-1}$ and the determinant $\chi$ in terms of $h^a_{\,\mu}$. As a result, the field strength $F^a_{\,\mu\nu}$ can be reduced into
\begin{eqnarray}
	F^a_{\,\mu\nu} &=& \partial_\mu \chi^a_{\,\nu} - \partial_\nu \chi^a_{\,\mu} = (\partial_\mu h^a_{\,\nu} - \partial_\nu h^a_{\,\mu})/2\,,
\end{eqnarray}
and the leading-order gravitational field equation in the GQFT is given by
\begin{widetext}
\begin{eqnarray}\label{eomGT}
	&&	\left[ \square h_a^{\,\rho} - \partial^\rho \partial_\nu h_a^{\,\nu} - \partial_\nu \partial_a h^{\nu \rho} + \partial_a \partial^\rho h + \delta^\rho_a (\partial_\nu\partial_\sigma h^{\nu \sigma} -\square h )  \right] + \gamma_W (\square h_a^{\,\rho}-\partial^\rho \partial_\nu h_a^{\,\nu})  = -16 \pi {G}_\kappa {\rm J}_a^{\,\rho}\,,
\end{eqnarray} 
\end{widetext}
where the $\gamma_W$-independent terms on the left-hand side (LHS) are derived from the LHS of Eq.~\eqref{EqGT} while the $\gamma_W$-dependent part from the term $m_G^2 {\cal D}_\nu \left( \chi \bar{\chi}_{aa^\prime}^{[\mu\nu]\mu^\prime\nu^\prime} \mathbf{F}^{a^\prime}_{\,\mu^\prime \nu^\prime} \right)$ contained in $J_a^{\,\mu}$. Here $\gamma_W \equiv \gamma_G (\alpha_G- \alpha_W/2)$ with $\gamma_G$ and $\alpha_{G(W)}$ defined in Ref.~\cite{Wu:2022mzr}. The current ${\rm J}_a^{\,\rho}$ on the right-hand side is composed of ordinary matter fields which sources the gravitational perturbation $h^a_{\,\mu}$. In Eq.~\eqref{eomGT} we have fixed the gauge conditions for the local ${\rm SP}(1,3)$ symmetry so that $h_a^{\,\mu}$ is a symmetric tensor with $h_a^{\,\mu} = h^\mu_{\,a}$. We also have $h\equiv \eta^\mu_a h^a_{\,\mu}$ and $h_{\mu\nu}\equiv \eta^a_\nu \eta_{\mu\rho} h^{\,\rho}_{a}$. However, the indices $a$ and $\mu$ in Eq.~\eqref{eomGT} do not possess any symmetry property, so that we can decompose this equation into the symmetric and anti-symmetric parts as follows
\begin{widetext}
\begin{eqnarray}\label{eomGTsym}
	\widetilde{G}_{\mu\nu}\equiv \frac{1}{2}\left[\square h_{\mu\nu} - 2\partial^\sigma \partial_{(\mu} h_{\nu)\sigma} + \partial_\mu\partial_\nu h + \eta_{\mu\nu} (\partial^\rho\partial^\sigma h_{\rho\sigma} -\square h ) \right] + \frac{\gamma_W}{2} [\square h_{\mu\nu} - \partial^\sigma \partial_{(\mu} h_{\nu)\sigma}]
	= -8\pi G_\kappa T_{(\mu\nu)}\,, 
\end{eqnarray} 
\end{widetext}
\begin{eqnarray}\label{eomGTanti}
	\widetilde{G}_{[\mu\nu]} \equiv -\frac{\gamma_W}{2} \partial^\sigma \partial_{[\mu} h_{\nu]\sigma} = -8\pi {G}_\kappa T_{[\mu\nu]}\,,
\end{eqnarray} 
where $\widetilde{G}_{\mu\nu}$ represents the generalized Einstein tensor and the source terms are given by
\begin{eqnarray}\label{TEMsa}
	T_{(\mu\nu)} &\equiv& (\eta_{\mu\rho} \eta^a_{\,\nu} + \eta_{\nu\rho}\eta^a_{\,\mu}) {\rm J}_a^{\,\rho}/2\,, \nonumber\\
	T_{[\mu\nu]} &\equiv& (\eta_{\mu\rho} \eta^a_{\,\nu} - \eta_{\nu\rho}\eta^a_{\,\mu}) {\rm J}_a^{\,\rho}/2 \,,
\end{eqnarray}
Eqs.~\eqref{eomGTsym} and \eqref{eomGTanti} comprises the complete linearized equations governing gravitational fields in the GQFT.

\section{Free Field Equations and Physical Degrees of Freedom}\label{SecFree}
As the first application of the general linearized field equations in the GQFT, this section is devoted to exploring the free equations in the absence of matter sources $T_{(\mu\nu)} = 0$ and $T_{[\mu\nu]} = 0$, paying attention to the question how many GW degrees of freedom (dofs) are propagating in this theory.

First of all, we follow the standard procedure in Refs.~\cite{Bertschinger:1993xt,Mukhanov:1990me,Nojiri:2020pqr} to decompose $h_{\mu\nu}$ into the following polarization modes:
\begin{itemize}
	\item Spin-2 tensor modes: $\hat{h}_{ij}$
	\begin{eqnarray}
		\hat{h}_{it} = \hat{h}_{tt} =0 \,,\, \hat{h}^i_i =0\,, \, \partial^i \hat{h}_{ij} = 0\,,
	\end{eqnarray}
	\item Spin-1 vector modes: $S_i$ and $F_i$
	\begin{eqnarray}
		h_{tt} = 0\,,\, h_{it} = S_i\,,\, h_{ij} = 2\partial_{(i}F_{j)}\,,\, \partial^i S_i = \partial^iF_i =0\,,
	\end{eqnarray}
	\item Spin-0 scalar modes: $\phi$, $B$, $\psi$ and $E$        
	\begin{eqnarray}
		h_{tt} = -2\phi\,, \, h_{it} = -\partial_i B,\, h_{ij} = -2\psi\delta_{ij} + 2 \partial_i \partial_j E.  
	\end{eqnarray}
\end{itemize}
We can conveniently summarize these fields with the following line element
\begin{eqnarray}\label{Metric}
  &ds^2&  = (1 + 2\phi) dt^2 - 2 (S_i -\partial_i B) dx^i dt  \nonumber\\
	& -& [\hat{h}_{ij} - (1-2\psi)\eta_{ij} + 2 \partial_{(i} F_{j)} + 2 \partial_i \partial_j E ]dx^i dx^j\,,
\end{eqnarray}
with $\eta_{ij} \equiv -\delta_{ij}$. Furthermore, one can prove that the field equations in Eqs.~\eqref{eomGTsym} and \eqref{eomGTanti} are invariant under the scalar-type gauge transformation $\delta h_{\mu\nu} = \partial_\mu\partial_\nu \zeta$, which leads the scalar fields $\phi$, $B$ and $E$ to transform as 
\begin{eqnarray}
	\phi \to \phi + \partial_t^2 \zeta /2\,, \quad B \to B+\partial_t \zeta\,,\quad E \to E-\zeta/2\,. \,\,
\end{eqnarray} 
Therefore, we can define the following two gauge-invariant variables
\begin{eqnarray}
	\Phi = \phi-\partial_t B/2\,\quad A = B+2\partial_t E\,.
\end{eqnarray} 

With the above polarization fields, the symmetric equation in Eq.~\eqref{eomGTsym} can give us the following independent equations 
\begin{itemize}
	\item $(t,t)$ component
	\begin{eqnarray}\label{EQtt}
		2\partial^k \partial_k \psi + \gamma_W \partial^k \partial_k \Phi = 0\,.
	\end{eqnarray}
	\item four-dimensional trace
	\begin{eqnarray}\label{EQtr}
		3\square  \psi =  \partial^i \partial_i \left( \psi + \Phi -  \partial_t A /2\right) \,.
	\end{eqnarray}
	\item $(t,i)$ component
	\begin{eqnarray}\label{EQti}
		&&	\left[\partial^k \partial_k (S_i -\partial_t F_i) + 4 \partial_t \partial_i \psi \right]  + (\gamma_W/2) \left[ \square S_i -  \square \partial_i A \right. \nonumber\\
		&+&   \left.  \partial^k \partial_k (S_i-\partial_t F_i) + 2 \partial_i\partial_t (\Phi-\psi+\partial_t A/2)  \right] 
		=0 \,.
	\end{eqnarray}
	\item $(i,j)$ component
	\begin{eqnarray}\label{EQij}
		&& (1+\gamma_W)\square \hat{h}_{ij} +  \gamma_W \square \partial_{(i} F_{j)} - {(\gamma_W+2)} \partial_t \partial_{(i} [S-\partial_t F]_{j)} \nonumber\\  
		&& + {2\eta_{ij} (\gamma_W+1)} \square \psi  \nonumber\\
		&&  + 2\partial_i \partial_j \left[ {(1-\gamma_W)}  \psi - \Phi +{(\gamma_W+1)}\partial_t A/2 \right] =0\,,
	\end{eqnarray}
\end{itemize}
It is clear that only the gauge-invariant fields appear in the final equations.

Our next task is to solve these field equations in the GQFT. We begin with the scalar sector. Eq.~\eqref{EQtt} implies the following constraint for $\psi$
\begin{eqnarray}\label{EqPsi}
	\psi = -\gamma_W \Phi/2 \,,
\end{eqnarray} 
which can be easily seen by transforming Eq.~\eqref{EQtt} into the Fourier space with a nonzero wave number. Moreover, we can obtain the additional independent relations 
\begin{eqnarray}
	\frac{\gamma_W}{2} \square A &=& \partial_t \left[ (4-\gamma_W) \psi +\gamma_W \Phi + \frac{\gamma_W}{2} \partial_t A \right]\,, \label{EQtiS} \\
  \square \psi &=& \partial_j \partial^j \left[ \frac{\gamma_W -1}{\gamma_W+1} \psi + \Phi - \frac{1}{2} \partial_t A \right]\,, \label{EQijS}
\end{eqnarray}
by acting one and two spatial derivatives on Eqs.~\eqref{EQti} and \eqref{EQij}, respectively. By solving Eqs.~\eqref{EqPsi}, \eqref{EQtr}, \eqref{EQtiS} and \eqref{EQijS}, we can yield the equation of motion for $\Phi$ 
\begin{eqnarray}
	\square \Phi = 0\,, \label{EqPhi} 
\end{eqnarray}  
and the constraint for $A$:
\begin{eqnarray}\label{Rela2}
	\partial_t A =  -(\gamma_W-2) \Phi \,.
\end{eqnarray}
 
We now turn to the vector sector containing $F_i$ and $S_i$. By taking into account Eqs.~\eqref{EqPhi}, \eqref{EqPsi}, and \eqref{Rela2}, all terms related to scalars are cancelled out in Eqs.~\eqref{EQti} and \eqref{EQij}, which results in the following reduced equations
\begin{eqnarray}
	&&\square S_i + \frac{(\gamma_W+2)}{\gamma_W} \partial_k \partial^k (S_i -\partial_t F_i) =0\,, \label{EQtiV}\\
	&& 	\square F_j - \frac{(\gamma_W+2)}{\gamma_W} \partial_t (S_j -\partial_t F_j) = 0\,, \label{EQijV}
\end{eqnarray} 
where the second equality follows by taking a divergence $\partial_i$ on Eq.~\eqref{EQij}. In order to fully shed light on the vector dynamics, we also need to consider the anti-symmetric gravitational equations in Eq.~\eqref{eomGTanti}, which can be written in terms of component fields as
\begin{itemize}
	\item $(t,i)$ component
	\begin{eqnarray}\label{ANTIti}
		\square S_i -\partial^j\partial_j (S_i - \partial_t F_i) = 0\,.
	\end{eqnarray}
	\item $(i,j)$ component
	\begin{eqnarray}\label{ANTIij}
		\square \partial_{[i} F_{j]} + \partial_t \partial_{[i}(S-\partial_t F)_{j]} = 0\,,
	\end{eqnarray}
\end{itemize}
where scalars are cancelled out totally due to their dynamics. By solving Eqs.~\eqref{EQtiV}-\eqref{ANTIij}, one can show for $\gamma_W\neq -1$
\begin{eqnarray}\label{EqVF}
	\square F_i  = 0\,,
\end{eqnarray} 
with the constraint
\begin{eqnarray}\label{EqVS}
	S_i = \partial_t F_i\,.
\end{eqnarray}

Finally, by considering the relations in the scalar and vector sectors, Eq.~\eqref{EQij} gives us
\begin{eqnarray}\label{EqT}
	\square \hat{h}_{ij} = 0\,,
\end{eqnarray}
which is the wave equation for the two massless tensor dofs like in the GR.

In summary, for a generic value of $\gamma_W \neq 0$ or $-1$, we have found five massless propagating GW dofs in the GQFT: two tensor modes $\hat{h}_{ij}$, two vector modes $F_i$, and one scalar $\Phi$, which can determine the dynamics of other component fields defined in Eq.~\eqref{Metric} via various relations.

\section{Newtonian Limits}\label{SecNew}

Let us now apply the linearized gravitational equations in the GQFT to explore the gravitational fields in the Newtonian limit. 
As is well known, the Newtonian limit~\cite{Carroll:2004st,Will:2018bme} is the situation in which the gravitational field is static and weak. 
Thus, in the following, all of the time derivatives in the linearized gravitational field equations can be ignored. Also, the matter is composed of dust and the configuration is static, so that the only nonzero component of the energy momentum tensor is only $T_{(00)} = \rho(\textbf{x})$, where the mass density $\rho$ is assumed to be a function of spatial coordinates $\textbf{x}$.  The antisymmetric energy-momentum tensor $T_{[\mu\nu]}$ vanishes by assumption.
Thus, the modified Einstein equations are given by
\begin{eqnarray}
	&&\widetilde{G}_{00}  = \partial_i \partial^i (2\psi + \beta \Phi) = -8\pi {G}_\kappa \rho({\bf x})\,, \nonumber\\
	&&\widetilde{G}_{0i}  = -\partial_k \partial^k \left[(1+\gamma_W) S_i -\gamma_W \partial_i A/2 \right]/2 = 0\,, \\
	&&\widetilde{G}_{ij} = -(1/2) \{ (1+\gamma_W) \partial_k \partial^k \hat{h}_{ij}  + \gamma_W \partial_k \partial_k \partial_{(i} F_{j)} \nonumber\\
	&&  - 2 \partial_i \partial_j [(\gamma_W-1)\psi + \Phi] - 2\eta_{ij} \partial_k \partial^k [(1-\gamma_W) \psi -\Phi] \} =0\,. \nonumber
\end{eqnarray}  
By solving these equations, the Newtonian potential $\Phi$ is shown to obey the conventional Poisson equation
\begin{eqnarray}\label{PhiN}
	\triangle \Phi \equiv - \partial_k \partial^k \Phi = 4\pi G_N \rho ({\bf x}) \,,
\end{eqnarray}
with the scalar $\psi$ determined by   
\begin{eqnarray}\label{PsiN}
	\psi = \Phi/(1-\gamma_W) \,,
\end{eqnarray}
while other fields $\hat{h}_{ij}$, $S_i$, $F_i$ and $A$ all vanish identically, where we have defined the measured Newtonian constant in terms of the fundamental coupling ${G}_\kappa$~\cite{Carroll:2004st,Will:2018bme} as follows
\begin{eqnarray}
	G_N \equiv \frac{1-\gamma_W}{(1-\gamma_W/2) (1+\gamma_W) } {G}_\kappa\,. 
\end{eqnarray}

One can probe this Newtonian field configuration with either non-relativistic test bodies or relativistic particles like photons. At low-energy limit of the GQFT, all these objects propagate along the geodesics described by:
\begin{eqnarray}\label{GeoEqn0}
	\frac{d^2 x^\rho}{d\tau^2} + \Gamma^\rho_{\, \mu\nu} \frac{d x^\mu}{d \tau} \frac{d x^\nu}{d\tau} = 0\,,
\end{eqnarray}   
where $\Gamma^\rho_{\,\mu\nu}$ is the usual Christoffel symbol defined with respect to the effective metric in Eq.~\eqref{Metric}. 

If the probing object is massive and moves very slowly, {\it i.e.}, $dx^i/d\tau \ll dt/d\tau$, the geodesic equation in Eq.~\eqref{GeoEqn0} dictates the object to follow the path
\begin{eqnarray}\label{NRacc}
	\frac{d^2 x^i}{dt^2} = \partial^i \Phi (\textbf{x}) = -\partial_i \Phi (\textbf{x})\,.
\end{eqnarray}
which is nothing but the acceleration of a massive body moving in the static Newtonian potential. Furthermore, Eq.~\eqref{NRacc} also implies that, regardless of the value of its mass, the test body would always experience the same acceleration and propagate with the same trajectory in this weak gravitational field configuration, which has been well tested by the experiments examining the Weak Equivalence Principle (WEP). Note that the best constraints on the WEP are provided by the E$\ddot{\rm o}$t-Wash group~\cite{Su:1994gu,Wagner:2012ui} and MICROSCOPE~\cite{MICROSCOPE:2022doy}. 

On the other hand, the gravitational field profile in the Newtonian limit can be detected by photons, which can lead to the gravitational light  deflection~\cite{Dyson:1920cwa,Shapiro:2004zz,Lambert:2009xy,lambert2011improved}, the Shapiro time delay~\cite{Shapiro:1964uw,Bertotti:2003rm} and the gravitational redshift~\cite{Pound:1960zz,Pound:1965zz,Vessot:1980zz} of photon frequencies. Note that these phenomena provide three of the most important tests of the GR in the history (see {\rm e.g.} Refs.~\cite{Will:2014kxa,Will:2018bme} for recent reviews and references therein). In what follows, we shall compute the relevant observables in the GQFT and use the current data to constrain the parameter $\gamma_W$. 

Since most observations for gravitational light deflections and time delays have been performed in the Solar system, we shall work in the following effective metric: 
\begin{eqnarray}\label{MetricNew}
	ds^2 = (1+2\Phi) dt^2 - (1-2\psi)\delta_{ij} dx^i dx^j\,,
\end{eqnarray}
where all other gravitational component fields vanish in the Newtonian limit according to previous discussions. Here we have chosen the gauge with $E=0$, so that $\phi = \Phi$, where $\Phi({\mathbf x})$ is the gauge-invariant Newtonian potential in the Solar system represented by
\begin{eqnarray}
	\Phi = -G_N M_\odot/r\,,
\end{eqnarray}
with $M_\odot$ denoting the solar mass and $r$ as the radial distance from the Sun. 
\begin{figure*}[ht!]
	\center
	\includegraphics[width=0.6 \linewidth]{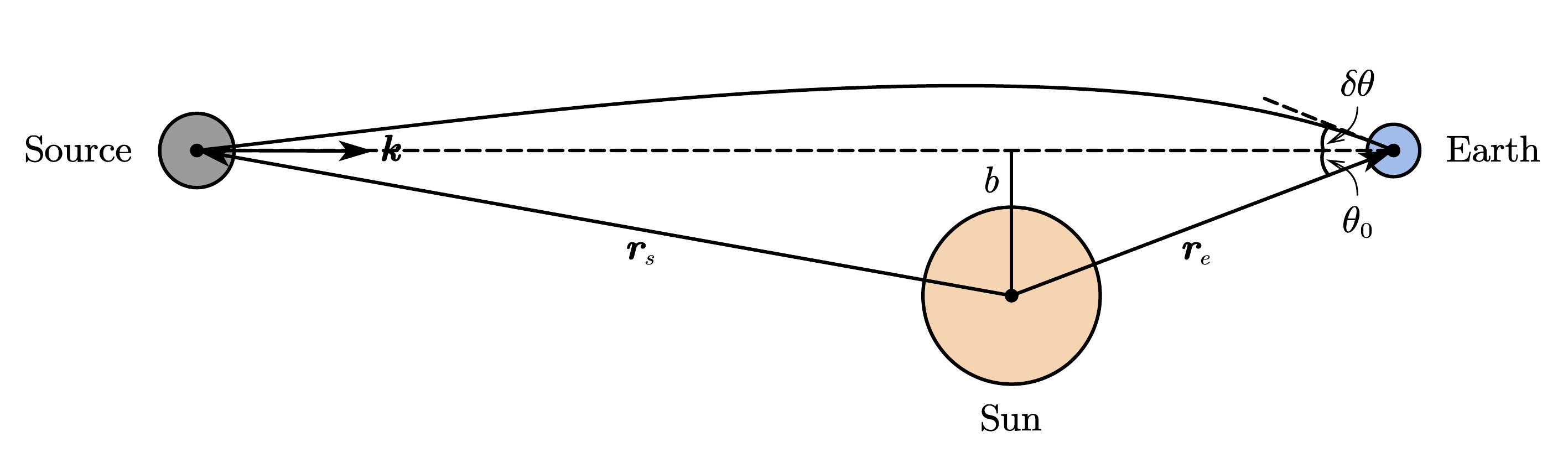}
	\caption{An illustration of the light ray deflection with the deflection angle $\delta \theta$ and the Shapiro time delay $\delta t_{\rm Shapiro}$ caused by the Newtonian field of a massive object, such as the Sun. }\label{FigLight}
\end{figure*}
Following the methods presented in Refs.~\cite{Will:2018bme,2014grav.book.....P}, we can derive, as illustrated in Fig.~\ref{FigLight}, the expressions for the deflection angle $\delta \theta$ and the Shapiro time delay $\delta t_{\rm Shapiro}$ as follows
\begin{eqnarray}\label{DefAngle}
	\delta \theta 
	&\approx& \left(2+\gamma_W\right) \left( {G_N M_\odot}/{b}  \right) \left(1+\cos\theta_0 \right) \,, \\
	\delta t_{\rm Shapiro} &\approx& 2 (2+\gamma_W) G_N M_\odot \ln \left({4 r_e r_s}/{b^2}\right)\,,
\end{eqnarray}
where $\theta_0$ is the elongation angle of the emitter relative to the Sun, $b$ is the closest approach from the Solar barycentric point to the line connecting the source and the Earth, and $r_{e(s)}$ is the radial distance of the Earth (source) from the Sun. Comparing with the well-known formulas in the parametrized post-Newtonian (PPN) formalism~\cite{Will:2014kxa,Will:2018bme}, it is seen that $\gamma_W$ is closely related to the PPN parameter $\gamma$ by $\gamma_W \approx \gamma-1$. 

Recent tremendous developments in the very-long-baseline radio interferometry (VLBI) and the radar time-delay experiments enable us to put strong constraints on the GQFT. In particular, the VLBI observations of the deflection angles of lights from quasars and radio galaxies have yielded $\gamma_W = (-0.8\pm 1.2)\times 10^{-4}$ at the $1\sigma$ CL~\cite{Lambert:2009xy,lambert2011improved}, while the most precise measurement of the Shapiro time delay is provided by the Cassini spacecraft~\cite{Bertotti:2003rm}, which has given the best limit to date on $\gamma_W = (2.1\pm 2.3)\times 10^{-5}$. 

Finally, we would like to discuss the gravitational redshift of photons~\cite{Pound:1960zz,Pound:1965zz,Vessot:1980zz} in the GQFT. Based on the arguments given in Ref.~\cite{Carroll:2004st}, we can derive the ratio of the photon frequencies $\omega({\bf x})$ at different locations ${\bf x}_1$ and ${\bf{x}_2}$:
\begin{eqnarray}
	\frac{\omega({\bf x}_2)}{\omega({\bf x}_1)} = \left(\frac{1+2\Phi({\bf x}_1)}{1+2 \Phi({\bf x}_2)}\right)^{1/2} \approx 1+ \Phi({\bf x}_1) - \Phi({\bf x}_2)\,,
\end{eqnarray}
where we have expanded the expression for $\Phi({\bf x}) \ll 1$. Note that the photon frequency modification only depends on the Newtonian potential $\Phi({\textbf x})$, without any reference to $\psi$, so that the gravitational redshift effect in the GQFT should be the same as in the GR.


\section{Conclusion and Discussion}\label{SecSum}
Understanding the nature of gravity and its quantization is one of main goals in the modern physics. Unlike the Einstein's GR which was based on the Riemann geometry, the GQFT~\cite{Wu:2022mzr}  has constructed the gravitational interaction based on the gauge principle that has been well-tested by other three fundamental interactions. In the present work, we have explored the fundamental physics and phenomenology in the weak gravity limit of the GQFT, so that the theory can be examined at the linear level of the perturbation $h_{\mu\nu}$. In order to realize this aim, we have derived the linearized gravitational field equations in the GQFT. It is found that, different from the usual diffeomorphism symmetry in the GR, the gauge symmetry in this theory is reduced to a scalar-type one parametrized by the infinitesimal gauge parameter $\zeta(x)$. Moreover, the difference between the GQFT and the GR at the linearized level can be parametrized by one single parameter $\gamma_W$. 

After establishing this linearized theory, we then apply this formalism to two special situations of important physical interest. In the first application, we examine the free linearized gravitational field equations in the absence of any matter fields. In particular, we focus on one crucial question:  how many and what physical propagating GW dofs are contained in this theory? As a result, different from the GR which includes only two massless tensor modes, there are five physical polarizations: two tensor modes, two vector modes and one scalar mode, all of which are massless. 

In the second application, we turn to the Newtonian limit in which the gravitational field is weak and the matter source fields are static. By solving the obtained field equations, we can obtain the conventional Poisson equation which connects the Newtonian potential with the matter density distribution. As a byproduct, we have obtained the exact relationship between the fundamental coupling ${G}_\kappa$ defined in the GQFT and the experimentally measured Newtonian constant $G_N$. We then make use of the non-relativistic objects and photons to probe the yielded gravitational field configuration. For a slowly-moving object, regardless of the value of its mass, it would always experience the same acceleration and follow the same trajectory in the gravitational field, which has been well tested by the experiments examining the WEP. Finally, we consider the motion of a photon in this Newtonian background, and investigate three classical tests: (i) the deflection of light, (ii) the time delay of light, and (iii) the gravitational  redshift,. It turns out that the GQFT gives exactly the same prediction of the gravitational redshift effect as in the GR, so that this kind of experiments cannot be used to distinguish these two theories. On the other hand, the light deflection and the Shapiro time delay do predict differently in the GQFT than in the GR, due to the dependence of the parameter $\gamma_W$. 
Thus, we can probe and constrain the GQFT with the associated experiments. In particular, the radar time-delay experiment carried out at the Cassini spacecraft provided the most stringent bound on $\gamma_W \lesssim {\cal O}(10^{-5})$.  

Besides the phenomena investigated in the present work, there are still many other aspects of the GQFT waiting us to explore. One classical test of the GR is provided by the perihelion advance of Mercury~\cite{Einstein:1915bz}, which requires the calculation of the gravitational field configuration beyond the linear level~\cite{Will:2018bme}. Furthermore, in the view of the additional propagating GW polarizations, some important questions are raised: How do the extra GW modes couple to matter fields? What are the corresponding coupling strengths? Is it possible to observe these GW dofs by the on-going and forthcoming GW observatories, such as LIGO-Virgo-KAGRA~\cite{LIGOScientific:2014pky,VIRGO:2014yos,Somiya:2011np,Aso:2013eba}, LISA~\cite{LISA:2017pwj}, Taiji~\cite{Hu:2017mde}, and TianQin~\cite{TianQin:2015yph}? We shall come back to these issues in the near future.

\section*{Acknowledgements}
\noindent This work was supported in part by the National Key Research and Development Program of China under Grant No.~2020YFC2201501, No.~2021YFC2203003, No.~2021YFC2202900, No.~2021YFC2201901, and No.~2017YFA0402204, and the National Natural Science Foundation of China (NSFC) under Grant No.~12005254, No.~12347103, No.~11825506, No.~11821505, and No.~12047503, and also the Strategic Priority Research Program of the Chinese Academy of Sciences as well as the Fundamental Research Funds for the Central Universities.


\bibliography{gwFp}

\end{document}